\documentstyle[psfig,onecolumn]{mn}

\begin{document}

\title{Weak lensing of the Sunyaev-Zel'dovich sky}

\author[Wu]{Xiang-Ping Wu\\
National Astronomical Observatories, Chinese Academy
                 of Sciences, Beijing 100012, China
}

\date{submitted 2003 July 11; accepted 2003 December 16 }        

\maketitle

\begin{abstract}
We address the question of whether the angular power spectrum of 
the thermal Sunyaev-Zel'dovich (SZ) sky  is further distorted 
by weak gravitational lensing of foreground large-scale structures.
Using an analytic approach to both gaseous and dark halo models,
we show that the contamination of weak lensing in the measurement of
SZ power power is negligibly small, and relatively corrections  
$|\Delta C_l|/C_l$ are less than $3\%$ up to $l=10^5$. This arises 
from both the weaker gravitational potentials of low-redshift 
matter inhomogeneities that can act as lenses for SZ sources (clusters) 
and the shallower shape of intrinsic SZ power spectrum at large $l$,
in contrast to the cosmic microwave background which can be 
significantly affected by weak lensing 
because of the distant location and significant damping of
its intrinsic power spectrum at small angular scales. 
 \end{abstract}

\begin{keywords}
cosmic microwave background ---
          galaxies: clusters: general --
          gravitational lensing ---
          intergalactic medium ---
          large-scale structure of universe
\end{keywords}

\vskip -3in           

\section{Introduction}

Gravitational lensing by foreground large-scale structures of 
the universe distorts many varieties of background sources.
Among these, detection of weak lensing-induced 
signatures in the cosmic microwave background (CMB) will provide 
a sensitive probe of matter distribution on large-scales
and at high redshifts up to $z\approx1100$ 
(e.g. Zaldarriaga \& Seljak 1999). On the other hand, 
the CMB photons can also be distorted by non-gravitational effect
along the path from the last scattering surface to us. 
The most significant pattern is the excess of the CMB power spectrum 
relative to primordial CMB anisotropy at $\ell>2000$ 
(Dawson et al. 2001; Bond et al. 2003),
which is produced by the inverse Compton scattering 
of the CBM photons by intervening energetic electrons confined 
within massive clusters, known as the Sunyaev-Zel'dovich (SZ) effect
(Sunyaev \& Zeldovich 1972). Studies of both the SZ-selected 
clusters and the SZ power spectrum yield important information
about the distribution and evolution of hot intracluster medium 
in the universe. 

At scales below 10 arcminutes characterized by $\ell\approx2000$--$10^4$, 
the CMB power spectrum is dominated by the thermal SZ effect 
of clusters distributed at intermediate/high 
redshifts up to $z\approx1$ (Cooray 2000; Cooray \& Hu 2000). 
An interesting question thus arises:  
Can the SZ sky be further distorted by the weak gravitational   
lensing effect of foreground matter inhomogeneities ?  Namely, 
we would like to address the question of 
to what extent the SZ power spectrum at small angular scales
should be corrected for because of weak lensing by
large-scale structures.  It has been noted that the deep gravitational
potentials of clusters can lead to both steplike wiggles on the CMB 
sky (Seljak \& Zaldarriaga 2000) and a systematic magnification bias 
of unresolved background radios behind clusters 
(Loeb \&  Refregier 1997).  
The latter may contaminate the measurement 
of the thermal SZ effect if the removal of background radio sources 
is made without inclusion of the inevitable lensing magnification 
bias.  Furthermore, the SZ sky is also correlated with the weak lensing 
map because the two effects are generated by common massive 
dark halos (Goldberg \&  Spergel 1999; Cooray \& Hu 2000).
The problem we explore in this paper is more 
straightforward, in the sense that the background source is now 
the SZ map which is associated with each cluster, 
while the foreground lenses are large-scale structures along the 
line of sight. Technically, we can use the formalism similar to the 
calculation of weak lensing of the CMB power spectrum, 
which has been well developed in the literature  
(Seljak 1996; Bernardeau 1997; Zaldarriaga \& Seljak 1999;
Hu 2000; Zaldarriaga 2000), except that the distribution function of 
background sources (clusters) should now be taken into account for 
our case. Throughout this paper we assume a flat cosmological
model ($\Lambda$CDM) with the parameters determined recently by 
{\sl WMAP}: $\Omega_{\rm M}=0.27$, $\Omega_{\Lambda}=0.73$, 
$\Omega_bh^2=0.0224$, $h=0.71$, $n_s=0.93$ and $\sigma_8=0.84$.

\section{Formalism}

Following the treatment of weak lensing of the CMB under the flat-sky
approximation (e.g. Bernardeau 1997; Zaldarriaga \& Seljak 1999;
Hu 2000; Zaldarriaga 2000), 
we calculate the lensed SZ temperature fluctuation 
$T(\mbox{\boldmath $\theta$})$
in the direction $\mbox{\boldmath $\theta$}$ by 
%
\begin{equation}
T(\mbox{\boldmath $\theta$})=T^s(\mbox{\boldmath $\theta$}+\nabla\phi)
            =\sum_{i=0}^{\infty}\frac{1}{i!}
            \left[\nabla\phi(\mbox{\boldmath $\theta$})\cdot \nabla\right]^i
	    	T^s(\mbox{\boldmath $\theta$}),
\end{equation}
where $T^s(\mbox{\boldmath $\theta$})$ is the unlensed SZ temperature
pattern in direction $\mbox{\boldmath $\theta$}$ produced by 
clusters at redshift $z_s$, $\nabla\phi$ denotes 
the lensing deflection angle by foreground large-scale structures 
along the line of sight, and the projected potential $\phi$ 
is related to the three-dimensional potential $\Phi({\bf r})$ 
of density perturbation $\delta({\bf r})$ at $z$ through
\begin{eqnarray}
\nabla\phi&=&\frac{2}{c^2}\int_0^{z_s} \left(\frac{D_{\rm ds}}{D_s}\right) 
             \nabla_{\bot}\Phi dr;\\
\nabla_{\bot}^2\Phi&=&\frac{3}{2}\Omega_MH_0^2\frac{\delta({\bf r})}{a},
\end{eqnarray}
in which $D_{\rm ds}$ and $D_s$ are the angular diameter distances from the
matter inhomogeneities at $z$ and the observer at $z=0$ to 
the sources at $z_s$, respectively, 
$r$ is the comoving distance and $a^{-1}=(1+z)$.
The power spectrum of the lensed SZ temperature map up to the fourth
order in deflection angle in terms of equation (1) is thus 
(e.g. Cooray 2003)
\begin{eqnarray}
C_l&=&C_l^s - C_l^s \int
              \frac{d^2{\bf l}^{\prime}}{(2\pi)^2}C^{\phi}_{l^{\prime}}
               \frac{({\bf l}\cdot{\bf l^{\prime}})^2}
                    {|{\bf l}^{\prime}|^4}  
         +\int\frac{d^2{\bf l}^{\prime}}{(2\pi)^2}C^{\phi}_{l^{\prime}}
                 C_{|{\bf l}-{\bf l}^{\prime}|}^s
              \frac{[({\bf l}-{\bf l}^{\prime})\cdot {\bf l}^{\prime}]^2}
                   {|{\bf l}^{\prime}|^4}  \nonumber \\
     & & +\frac{1}{2} C_l^s \left[\int \frac{d^2{\bf l}^{\prime}}{(2\pi)^2}
                C^{\phi}_{l^{\prime}} \frac{({\bf l}\cdot{\bf l^{\prime}})^2}
                    {|{\bf l}^{\prime}|^4} \right]^2 
         - \int \frac{d^2{\bf l}^{\prime}}{(2\pi)^2}
                \frac{d^2{\bf l}^{\prime\prime}}{(2\pi)^2}
                C_{|{\bf l}-{\bf l}^{\prime}|}^s
                C^{\phi}_{l^{\prime}} C^{\phi}_{l^{\prime\prime}}
                 \frac{[({\bf l}-{\bf l}^{\prime})\cdot {\bf l}^{\prime}]^2}
                   {|{\bf l}^{\prime}|^4} 
                 \frac{[({\bf l}-{\bf l}^{\prime})\cdot 
                    {\bf l}^{\prime\prime}]^2}{|{\bf l}^{\prime\prime}|^4}
                                    \nonumber \\
     & & + \frac{1}{2}\int \frac{d^2{\bf l}^{\prime}}{(2\pi)^2}
                \frac{d^2{\bf l}^{\prime\prime}}{(2\pi)^2}
                C_{|{\bf l}-{\bf l}^{\prime}-{\bf l}^{\prime\prime}|}^s
                 C^{\phi}_{l^{\prime}} C^{\phi}_{l^{\prime\prime}}
                 \frac{[({\bf l}-{\bf l}^{\prime}-{\bf l}^{\prime\prime})
                      \cdot {\bf l}^{\prime}]^2}
                   {|{\bf l}^{\prime}|^4} 
                 \frac{[({\bf l}-{\bf l}^{\prime}-{\bf l}^{\prime\prime})
                      \cdot {\bf l}^{\prime\prime}]^2}
                     {|{\bf l}^{\prime\prime}|^4}
\end{eqnarray}
where $C_l^s$ is the unlensed power spectrum of the SZ map, and 
$C_l^{\phi}$ corresponds to the lensing power spectrum defined as
\begin{equation}
C_l^{\phi}=\frac{9\Omega^2_{\rm M}H_0^4}{c^4}
            \int_0^{z_{\rm dec}}w(z_s)dz_s\int_0^{z_s}
             \left(\frac{D_{\rm ds}}{D_s}\right)^2\frac{dr}{a^2}
             P^{\rm m}\left(\frac{l}{D_A},z\right),
\end{equation}
$z_{\rm dec}\approx1100$ is the decoupling redshift, $D_A$ is the 
comoving angular diameter distance,  $P^{\rm m}(k,z)$ is the matter power
spectrum at $z$, and $w(z_s)$ is the distribution function of 
the SZ sources (i.e. clusters).  If the mass function of dark halos is
$d^2n/dMdV$, then $w(z_s)$ can be expressed as
\begin{equation}
w(z_s)=\frac{ \frac{d^2V}{dz_sd\Omega}
              \int_{M_{\rm min}}^{\infty} dM\;
                  \frac{d^2n(M,z_s)}{dMdV} }
            {\int_0^{z_s}\frac{d^2V}{dz_sd\Omega}dz_s
              \int_{M_{\rm min}}^{\infty} dM\;
                  \frac{d^2n(M,z_s)}{dMdV} },
\end{equation}
where $d^2V/dz_sd\Omega$ denotes the comoving volume per unit 
redshift and per steradian. 
We have also tested the observationally determined
X-ray luminosity function of clusters 
(Rosati, Borgani \& Norman 2002)  for $L_X>10^{43}$ erg s$^{-1}$
instead of the mass function 
$d^2n/dMdV$ and found that the modification is only minor.  
In principle, the lower mass limit $M_{\rm min}$ can be taken 
to be 0 because low-mass halos (e.g. $M<10^{13}M_{\odot}$) almost
make no contributions to SZ effect unless the self-similar model
is adopted for gas distribution. Nevertheless, 
inclusion of the low-mass halos
may significantly alter the distribution function $w(z_s)$ . 
So, we will take $M_{\rm min}=10^{14}$ $M_{\odot}$  
in the following calculation, which corresponds roughly to a lower
luminosity cutoff of $L_X>10^{43}$ erg s$^{-1}$.
Such a choice of the lower limit 
is also consistent with the prediction of preheating model, 
in which the preheated baryons cannot be trapped in the shallower 
gravitational potential wells of low-mass halos (see below).

The undistorted thermal SZ power spectrum $C_l^s$ is composed of the 
so-called Poisson term $C_l^P$ and clustering term $C_l^C$,
respectively, 
\begin{eqnarray}
C_l^P &=& g^2(x)\int_0^{z_{\rm dec}} dz_s\frac{d^2V}{dz_sd\Omega}
        \int_{M_{\rm min}}^{\infty} 
	dM\frac{d^2n(M,z_s)}{dMdV}   \vert y_l(M,z_s)\vert ^2; \\
C_l^C &=& g^2(x)\int_0^{z_{\rm dec}} dz_s\frac{d^2V}{dzd\Omega}
                 P^{\rm lin}\left(\frac{l}{D_0},z_s\right) 
          \left[\int_{M_{\rm min}}^{\infty} dM
         \frac{d^2n(M,z_s)}{dMdV}b(M,z_s)\vert y_l(M,z_s)\vert\right]^2,
\end{eqnarray}
where  $y_l(M,z_s)$ is the Fourier transform of 
the Compton $y$-parameter of the cluster of mass $M$ at $z_s$,  
$D_0=(1+z_s)D_s$ is the comoving distance, 
$P^{\rm lin}(k,z_s)$ is the linear matter power spectrum
at $z_s$, and 
$b(M,z_s)$ is the bias parameter, for which we use
the analytic prescription of (Mo \& White 1996). The frequency
dependence term $g(x)$ is defined as 
$g(x)=(x^2 e^x)[4-x \coth(x/2)]/(e^x-1)^2$,
in which $x=h\nu/kT_{\rm CMB}$ is the dimensionless frequency, and 
$T_{\rm CMB}$ is the temperature of the present CMB. The observing 
frequency will be fixed at $\nu=30$ GHz in this study.

Mass function of massive dark halos is assumed to follow 
the standard Press-Schechter function
\begin{equation}
\frac{d^2n(M,z)}{dMdV}=-\sqrt{\frac{2}{\pi}} \frac{\bar{\rho}}{M} 
    \frac{\delta_{\rm c}}{\sigma^2} 
    \frac{d\sigma}{dM} 
    \exp{\left(-\frac{\delta_c^2}{2\sigma^2} \right)},
\end{equation} 
in which $\sigma$ is the linear theory variance of the mass density
fluctuation in sphere of mass $M$:
$\sigma^2=(1/2\pi^2) \int_0^{\infty} 
k^2 P^{\rm lin}(k){\vert W(kR)\vert}^2 dk$,  
$W(kR)$ is the Fourier representation of the window function, and 
$W(kR)=3([\sin(kR)-kR\cos (kR)]/(kR)^3$. We adopt the present-day linear 
matter power spectrum $P^{\rm lin}(k)$ given by Bardeen et al. (1986) 
and a power-law form ($\propto k^{n_s}$) 
for the primordial matter density fluctuation.  

At small scales below $\sim10^{\prime}$, matter power 
spectrum is strongly affected by non-linear 
structures. Following Seljak (2000) and Cooray, Hu \& Miralda-Escud\'e (2000)
we use a halo approach to calculate the non-linear dark matter
power spectrum $P^{\rm m}(k,z)$, which consists of 
a single halo term $P^{\rm 1h}(k,z)$ and a clustering term
$P^{\rm 2h}(k,z)$:
\begin{eqnarray}
P^{\rm 1h}(k,z)&=&\int dM \frac{d^2n(M,z)}{dMdV} 
                  \left|\frac{\rho_h(M,z,k)}
                             {\overline{\rho}(z)}\right|^2 \nonumber\\
P^{\rm 2h}(k,z)&=&P^{\rm lin}(k,z)  
                   \left[\int dM \frac{d^2n(M,z)}{dMdV} b(M,z)
                     \frac{\rho_h(M,z,k)}{\overline{\rho}(z)}\right]^2,
\end{eqnarray}
where $\rho_h(M,z,k)$ is the Fourier transform of dark matter density
profile $\rho_h(M,z)$ of a halo $M$ at $z$, and $\overline{\rho}(z)$
is the mean mass density of the background universe at $z$.  
We adopt the functional form suggested by
numerical simulations (Navarro, Frenk \& White 1997; NFW)
for $\rho_h(M,z)$ and fix the 
concentration parameter in terms of the empirical fitting formula 
of  Bullock et al. (2001). 

Finally, we need to specify the gas distribution of dark halos 
in order to evaluate the Compton $y$-parameter. Since the weak
lensing effect of the SZ map may occur at very small angular scales, 
we should employ a more realistic model for the gas distribution in the 
central regions of clusters instead of the commonly adopted self-similar 
model in literature. Recall that the SZ power
spectra predicted by the self-similar model and other 
physically-motivated models 
such as preheating and radiative cooling become
indistinguishable roughly at $l<2000$. For simplicity, here we work with 
an analytic preheating model, in which the gas distribution in 
massive halos can be obtained by combining the equation of hydrostatic 
equilibrium with a modified entropy profile 
$S(r)=S_{\rm floor}+S_{\rm NFW}(r)$. Namely,  
a constant entropy floor of $S_{\rm floor}\approx120$ keV cm$^2$  
is now added to the entropy 
profile $S_{\rm NFW}(r)$ predicted by the self-similar model which again
takes the universal density profile (NFW) .
This model allows us to essentially reproduce the observed X-ray surface 
brightness distribution (i.e. $\beta$ model) and statistical 
properties of groups and clusters (e.g. the X-ray Luminosity -
temperature relation and entropy distribution)
(Babul et al. 2002; Xue \& Wu 2003).

\section{Results}

We first demonstrate in Figure 1 the lensing power spectrum $C_l^{\phi}$
for the SZ sky. For comparison,  we have also shown the result 
for the CMB calculated by setting $w(z_s)=\delta(z_s-z_{\rm dec})$ in 
eqn.(5) and running the publicly available code CMBFAST. 
At small scales, the lensing  power spectra 
for the SZ and CMB maps appear to be similar in shape, but the 
lensing amplitude for the SZ sources is about two orders of magnitude 
smaller than that for the CMB. This is simply because there are
much more intervening matter inhomogeneities acting as lenses for 
the CMB photons than those for the SZ signals generated by
clusters located mostly at $z<1$.  Furthermore, 
$C_l^{\phi}$ peaks at lower $l$ for the SZ map than for the CMB,
consistent with the fact that the large-scale structures responsible 
for weak lensing of the SZ sky are at low redshifts and therefore,
subtend relatively large angles.  Recall that gravitational 
lensing conserves the
surface brightness of background source. Therefore, $C_l^{\phi}$ 
acts as a narrow window function on the background power spectrum
$C_l^s$ via eqn.(4), which alters the intrinsic patterns of $C_l^s$.
Our intuition in terms of 
this simple argument is that the weak lensing effect on the SZ
power spectrum is probably much smaller than the lensing signature 
on the CMB, if the lensing induced correction is simply 
proportional to $C_l^{\phi}$. Finally,  the lensing  power spectra 
at small scales are dominated by non-linear structures, and 
linear approximation of matter power spectrum $P^{\rm lin}(k,z)$
may gives rise to an underestimate of the lensing effect.

\begin{figure} 
	\psfig{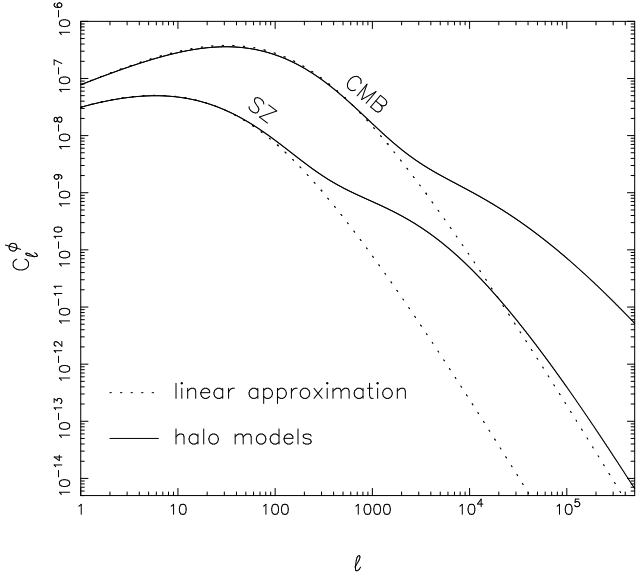}
	\caption{Power spectra of weak lensing for CMB and SZ map.
Dotted and solid lines represent the results of linear and non-linear 
matter power spectra, respectively.}
   \end{figure}

Nevertheless, the extent to which  background sources are affected
by weak lensing of foreground large-scale structures also depends 
critically on the shape of the source power spectrum.
We begin by recalling how the CMB can effectively be lensed 
by large-scale structures (see Hu 2000): At large scales 
$l<100$, the CMB power spectrum  can be considered
to be a slowing varying function of $l$ (see Figure.2), and 
we can approximately take $C_l^s$ out of the second integral in eqn(4). 
The lensing induced two terms to the first order approximation
in deflection angle cancel if ${\bf l}-{\bf l}^{\prime}$ 
is further replaced by ${\bf l}$. This yields $C_l\approx C_l^s$.
At small scales, existence of the strong Silk damping of 
intrinsic CMB power spectrum at $l>2000$ makes the contribution
of the first correction term in eqn.(4) considerably small,
and the second term dominates the lensing correction.  
As a result, weak lensing always leads to an increase of the 
CMB power at small angular scales, as shown in Figure 2.
Now we turn to the SZ power spectrum: Unlike the intrinsic 
CMB, there is no significant damping effect on the thermal 
SZ power spectrum over all angular scales. Rather, the SZ
power spectrum has a long, slowly decreasing tail at high $l$. 
In other word, the first negative term in eqn.(4) can no 
longer be ignored. Applying the approximation that 
$C_l^s$ is slowly varying and 
${\bf l}-{\bf l}^{\prime}\approx {\bf l}$ to eqn.(4), we have 
$C_l\approx C_l^s$ to the first order in deflection angle. 
Our numerical calculation reaches 
essentially the same conclusion. In Table 1 we have listed 
the relative corrections $|C_l-C_l^s|/C_l^s$ for 
a set of multipoles from $l=100$ to $l=10^5$. It turns out
that the lensing induced corrections are well below $3\%$ even up 
to $l\approx10^5$. Meanwhile, we have given in Table 1 the 
power spectrum ratios of the fourth order correction 
[the last three terms of eqn.(4)] to the second order approximation 
in $C_l^{\phi}$ [the second and third terms of eqn.(4)]. 
While the contribution of higher order corrections 
to the power spectrum $C_l$ becomes increasingly important with
the decrease of angular scale, the effect of higher order terms
on the evaluation of $C_l$ is still negligibly small on scales 
up to $l=10^5$ as compared with the contribution from the first 
order approximation in deflection angle.

\begin{table*}
\vskip 0.2truein
\begin{center}
\caption{Numerical Results}
\vskip 0.2truein
\begin{tabular}{ccccc}
\hline
$l$  &  100  & 1000 & 10000  & 100000\\
\hline
$\frac{|C_l-C_l^s|}{C_l^s}\;(\%)$ & 0.0025 & 0.0065 & 0.53  &  2.51\\
$\frac{C_l({\rm fourth\; order})}{C_l({\rm second\; order})}$ &
        $7.9\times10^{-10}$ &  $9.2\times10^{-9}$ &
        $1.5\times10^{-5}$  &  $6.7\times10^{-4}$ \\
\hline
\end{tabular}
\end{center}
 \end{table*}

\begin{figure} 
	\psfig{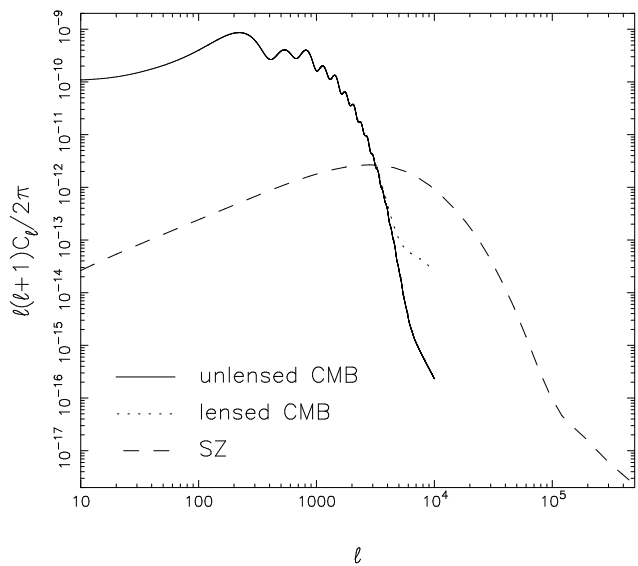}
	\caption{Angular power spectra of the unlensed and lensed CMB and
SZ map. The lensed SZ power spectrum is almost identical to the 
intrinsic one, and the relative difference reaches only $3\%$ at 
$l\approx10^5$.}
   \end{figure}

\section{Conclusions}

It has been shown that the angular power spectrum of the thermal SZ
sky is almost unaffected by the weak lensing of foreground large-scale 
structures. This is primarily due to both the relatively weaker 
gravitational potentials of low-redshift matter inhomogeneities 
that can act as lenses for SZ sources and the shallower shape of 
SZ power spectrum at smaller angular scales. Note that we have adopted
the gas density profile predicted by the so-called preheating
model, which results in a more rapid decline of 
the SZ power spectrum at large $l$ than the one predicted by
the self-similar model or cooling model (Wu \& Xue 2003). In other
word, we have already provided an optimistic estimate of the weak
lensing effect on the SZ power spectrum since significant 
fluctuations intrinsic to background sources are more easily 
be distorted by lensing. This explains why weak lensing of
large-scale structures of the universe can effectively
alter the CMB sky at $z_s\approx 1100$ as the CMB power
spectrum exhibits the sharp damping signatures at high $l$ 
(Hu 2000). The angular power spectrum of 21cm map of the reionization
at $z\sim10$ also demonstrates a significant decline feature at 
$l=10^3-10^4$ (e.g. Zaldarriaga, Furlanetto \& Hernquist 2003),
and it is thus expected that the 21cm fluctuation of the reionization
should be affected by weak lensing (Pen 2003).
Nonetheless, the present study has eliminated the concern that the SZ 
sky may be further contaminated by weak lensing of
foreground large-scale structures of the universe.

\section*{Acknowledgements}

This work was supported by
the National Science Foundation of China, under Grant No. 19725311
and the Ministry of Science and Technology of China, under Grant
No. NKBRSF G19990754.


\begin{thebibliography}{}

\bibitem{} Babul, A., Balogh, M. L., Lewis, G. F.,  \& Poole, G. B.,
 	      2002, MNRAS,  330, 329.

\bibitem{} Bardeen, J. M., Bond, J. R., Kaiser, N., \& 
		Szalay, A. S., 1986, ApJ, 304, 15.

\bibitem{} Bernardeau, F.,  1997, A\&A, 324, 15.

\bibitem{} Bond,  J. R., et al., 2003, ApJ, in press 
		(astro-ph/0205386).

\bibitem{} Bullock, J.~S., et al., 2001, MNRAS, 321, 559.

\bibitem{} Cooray, A., 2000, PRD, 62, 103506.

\bibitem{} Cooray, A., 2003, astro-ph/0309301

\bibitem{} Cooray, A., \& Hu, W., 2000, ApJ, 534, 533.

\bibitem{} Cooray, A., Hu, W., \& Miralda-Escud\'e, J., 2000,  
	       ApJ, 535, L9.

\bibitem{} Dawson, K. S., et al., 2001, ApJ, 553, L1.

\bibitem{} Goldberg, D. M., \& Spergel, D. N., 
		1999, PRD, 59, 10300.

\bibitem{} Hu, W., 2000, PRD, 62, 043007.

\bibitem{} Loeb, A., \& Refregier, A.,  1997, ApJ, 476, L59.

\bibitem{} Mo, H.-J., \& White, S. D. M., 1996, MNRAS, 282, 347.

\bibitem{} Navarro, J. F., Frenk, C. S., \& White, S. D. M., 1997, MNRAS, 
	    	490, 493.

\bibitem{} Pen, U.-L., 2003, PRD, submitted (astro-ph/0305387).

\bibitem{} Rosati, P., Borgani, S., \& Norman, C., 2002, ARAA, 40, 539.

\bibitem{} Seljak, U., 1996, ApJ, 463, 1.

\bibitem{} Seljak, U., 2000, MNRAS, 318, 203.

\bibitem{} Seljak, U., \&  Zaldarriaga, M., 2000, ApJ,  538, 57.

\bibitem{} Sunyaev, R. A., \& Zeldovich, Y. B., 1972, 
		Comm. Astrophys. Space Phys., 4, 173.

\bibitem{} Wu, X,-P., \&  Xue, Y.-J., 2003, ApJ, 590, 8.

\bibitem{} Xue, Y.-J., \& Wu, X.-P., 2003, ApJ, 584, 34.

\bibitem{} Zaldarriaga, M., 2000,  PRD,  62, 063510.

\bibitem{} Zaldarriaga, M., Furlanetto, S. R., \& Hernquist, L., 2003,
		ApJ, submitted (astro-ph/0311514)

\bibitem{} Zaldarriaga, M., \& Seljak, U., 1999, PRD, 59, 123507.


\end{thebibliography}
\end{document}